\documentstyle[12pt,epsf]{article}
\def \beq{\begin{equation}}
\def \eeq{\end{equation}}
\begin{document}
\rightline{TECHNION-PH-95-28}
\rightline{EFI-95-66}
\rightline{hep-ph/9510363}
\rightline{October 1995}
\bigskip
\bigskip
\centerline{{\bf DETERMINING THE CKM UNITARITY TRIANGLE FROM}}
\centerline{{\bf $B$ DECAYS TO
CHARGED PIONS AND KAONS}\footnote{To be submitted to Physical Review Letters.}}
\bigskip
\centerline{\it Michael Gronau}
\centerline{\it Department of Physics}
\centerline{\it Technion -- Israel Institute of Technology, Haifa 32000,
Israel}
\medskip
\centerline{and}
\medskip
\centerline{\it Jonathan L. Rosner}
\medskip
\centerline{\it Theoretical Physics Division, Fermi National Accelerator
Laboratory}
\centerline{\it P. O. Box 500, Batavia, IL 60510}
\smallskip
\centerline{and}
\smallskip
\centerline{\it Enrico Fermi Institute and Department of Physics}
\centerline{\it University of Chicago, Chicago, IL 60637
\footnote{Permanent address.}}
\bigskip
\centerline{\bf ABSTRACT}
\medskip
\begin{quote}
Decay rates of $B^0(t)\to\pi^+\pi^-,~B^0\to\pi^- K^+,~B^+\to\pi^+ K^0~
(K_S\to\pi^+\pi^-$) and of charge-conjugate processes are studied within flavor
SU(3) symmetry and first-order SU(3) breaking. We show that these measurements
can determine with a reasonable accuracy the two angles, $\alpha$ and $\gamma$,
of the Cabibbo-Kobayashi-Maskawa unitarity triangle.
\end{quote}
\bigskip

$B$ decays provide a variety of CP asymmetry measurements \cite{CPreview},
which can test the currently favored hypothesis that phases in elements of the
Cabibbo-Kobayashi-Maskawa (CKM) matrix \cite{CKM} are the source of the
observed CP violation in the neutral kaon system \cite{CCFT}. The
time-dependent rate-asymmetry between the process $B^0(t)\to\pi^+\pi^-$ and its
CP-conjugate measures one of these phases, the angle $\alpha$ of the CKM
unitarity triangle. Penguin amplitudes \cite{Penguin} and higher order
electroweak penguin contributions \cite{EWP} complicate the situation somewhat.
However, by measuring also the rates of
$B^0\to\pi^0\pi^0,~B^+\to\pi^+\pi^0$
and of their charge-conjugate counterparts one can isolate the amplitudes
contributing to final states with isospin $0$ and $2$ and thereby determine
$\alpha$ with a rather good accuracy \cite{Isospin,EWPin}. The detection of the
modes involving neutral pions poses an interesting challenge for future
experiments.

A few alternative ways to learn the penguin effect in $B^0\to\pi^+\pi^-$ were
suggested recently.
DeJongh and Sphicas \cite{DS} have studied in detail the dependence of the
asymmetry in $B^0(t)\to\pi^+\pi^-$ on the (unknown) magnitude and relative
phase
of the tree and penguin amplitudes contributing to this process.
Using flavor SU(3) symmetry, Silva and Wolfenstein \cite{SilWo}
proposed to approximately estimate the penguin contribution by comparing the
tree-dominated decay rate of $B^0\to\pi^+\pi^-$ with that of $B^0\to\pi^- K^+$
which has a large penguin term. Buras and Fleischer \cite{BurFl} suggested to
isolate the penguin term
in $B^0\to\pi^+\pi^-$ from its (SU(3)-related) dominant effect in
the time-dependent asymmetry of $B^0(t)\to K^0\overline{K}^0$.

In the present report we describe a method which determines simultaneously both
the angle $\alpha$ and the angle $\gamma$ of the unitarity triangle from the
decay rates of $B^0(t)\to\pi^+\pi^-,~B^0\to\pi^- K^+,~B^+\to\pi^+ K^0$ (where
$K^0\to K_S\to\pi^+\pi^-$) and their charge-conjugates. All these modes are
detected by charged pions and kaons in the final state. Other ways to measure
$\gamma$, based on charged $B$ decays, were proposed in Ref.~\cite{gamma}. Our
method employs flavor SU(3) symmetry \cite{DZ,SW,Chau}, and neglects
``annihilation'' amplitudes in which the spectator quark (the light quark
accompanying the $b$ in the initial meson) enters into the decay Hamiltonian
\cite{SU3}. These amplitudes in $B$ decays are expected to be suppressed by
$f_B/m_B$, where $f_B \simeq 180$ MeV. In order to improve the precision of the
method, we also include first-order SU(3) breaking terms \cite{SU3br}. Second
order corrections, which are expected to be at a level of a few percent, will
be neglected.

In the SU(3) limit and neglecting annihilation terms all $B$ decay amplitudes
into $\pi\pi~,\pi K$ and $K\overline{K}$ states can be decomposed in terms of
three independent amplitudes \cite{EWPin,SU3}:  a ``tree'' contribution
$t(t')$, a ``color-suppressed'' contribution $c(c')$ and a ``penguin''
contribution $p(p')$. These amplitudes contain both the leading-order and
electroweak penguin contributions:
\beq
t \equiv T + (c_u - c_d) P_{EW}^C~~,~~~c \equiv C + (c_u - c_d) P_{EW}~~,
{}~~~p \equiv P + c_d P_{EW}^C~~.
\eeq
Here the capital letters denote the leading-order contributions defined in
Ref.~\cite{SU3}, and $P_{EW}$ and $P_{EW}^C$ are color-favored and
color-suppressed electroweak penguin amplitudes defined in Ref.~\cite{EWPin}.
The values $c_u = 2/3$ and $c_d = -1/3$ are those which would follow if
the electroweak penguin coupled to quarks in a manner proportional to their
charges.  (Small corrections, which we shall ignore
and which do not affect our analysis,
arise from axial-vector $Z$ couplings and from $WW$
box diagrams.)  The $\Delta S = 0$ amplitudes are denoted by unprimed
quantities and the $\Delta S = 1$ processes by primed quantities.

The amplitudes of the two processes $B^0\to\pi^+\pi^-$ and $B^0\to\pi^-
K^+$ are expressed as
$$
A_{\pi\pi}\equiv A(B^0\to\pi^+\pi^-)=-t-p=-T-P-{2\over 3}P^C_{EW}~~,
$$
\beq \label{eqn:amps}
A_{\pi K}\equiv A(B^0\to\pi^- K^+)=-t'-p'=-T'-P'-{2\over 3}P'^C_{EW}~~,
\eeq
while that for $B^+\to \pi^+ K^0$ will be approximated by
\beq \label{eqn:approx}
A_+\equiv A(B^+\to\pi^+ K^0)= p' = P'-{1\over 3}P'^C_{EW}
\approx P' + {2\over 3}P'^C_{EW}~~,
\eeq
neglecting a color-suppressed electroweak penguin effect of order
$|P'^C_{EW}/P'|={\cal O}((1/5)^2)$ \cite{EWPin}. With this approximation,
$A_+$ contains the same combination of electroweak and gluonic penguins
as in the expression for $A_{\pi K}$.

The terms on the right-hand-sides of (\ref{eqn:amps}) and (\ref{eqn:approx})
carry well-defined weak phases. The
weak phase of $T$ is ${\rm Arg}(V_{ud}V^*_{ub})=\gamma$, and that of $P+{2\over
3}P^C_{EW}$ is approximately ${\rm Arg}(V_{td}V^*_{tb})=-\beta$, where we
neglect corrections due to quarks other than the top quark. The effects of the
$u$ and $c$ quarks become appreciable \cite{BF} when $V_{td}$ obtains its
currently allowed smallest values. This corresponds to a small deviation of the
CP asymmetry in $B^0(t)\to\pi^+\pi^-$ from $\sin(2\alpha)\sin(\Delta mt)$
(where $\Delta m$ is the neutral $B$ mass-difference). For large values of
$V_{td}$, where the deviation due to the penguin amplitude
becomes significant \cite{MG}, the $u$ and $c$
contributions become very small. $T'$ also carries the phase $\gamma$, while
the weak phase of $P'+{2\over 3}P'^C_{EW}$ is ${\rm Arg}(V_{ts}V^*_{tb})=\pi$.
The ratio of $\Delta S = 1$ to $\Delta S = 0$ tree and penguin amplitudes are
given by the corresponding ratios of CKM factors, $|T'/T|=|V_{us}/V_{ud}|
\equiv r_u=0.23,~|P'/P|=|V_{ts}/V_{td}|\equiv r_t$.

Denoting ${\cal T}\equiv |T|,~{\cal P}\equiv |P+{2\over 3}P^C_{EW}|$ and
assigning SU(3)-symmetric strong phases $\delta_T,~\delta_P$ to terms with
specific weak phases, (\ref{eqn:amps}) and (\ref{eqn:approx}) may be
transcribed
 as
$$
A_{\pi\pi}={\cal T}e^{i\delta_T}e^{i\gamma}+{\cal P}e^{i\delta_P}e^{-i\beta}~~,
$$
$$
A_{\pi K}=r_u{\cal T}e^{i\delta_T}e^{i\gamma}-r_t{\cal P}e^{i\delta_P}~~,
$$
\beq \label{eqn:amprev}
A_+=r_t{\cal P}e^{i\delta_P}~~.
\eeq

To introduce first-order SU(3) breaking corrections, we note that in the $T'$
amplitude the $W$ turns into an $\overline{s}$ quark instead of a
$\overline{d}$ in $T$. This SU(3) breaking term was denoted by $T'_1$ in
Ref.~\cite{SU3br}. Assuming factorization for $T$, which is supported by
experiments \cite{Browder}, SU(3) breaking is given by the $K/\pi$ ratio of
decay constants
\beq
{{\cal T'}\over {\cal T}}={|V_{us}|\over |V_{ud}|}{f_K\over f_{\pi}}\equiv
\tilde{r}_u~~.
\eeq
In the penguin amplitudes (including electroweak penguin) of both $B^0\to\pi^-
K^+$ and $B^+\to\pi^+ K^0$ the $\overline{b}$  quark turns into an
$\overline{s}$ quark instead of a $\overline{d}$ in $B^0\to\pi^+\pi^-$. This
SU(3) breaking term was denoted by $P'_1$ in Ref.~\cite{SU3br}. Here we will
denote the magnitude of the $\Delta S = 1$ penguin amplitude by $r_t\tilde{\cal
P}$, to allow for SU(3) breaking. Since factorization is questionable for
penguin amplitudes, one generally expects $\tilde{\cal P}\ne(f_K/f_{\pi}){\cal
P}$. We will assume that the phase $\delta_P$ is unaffected by SU(3) breaking.
Since this phase is likely to be small \cite{Penphase}, this assumption is not
expected to introduce  a significant uncertaintly in the determination of the
weak phases.

Thus, including first-order SU(3) breaking,
Eqs.~(\ref{eqn:amprev}) are modified to become
$$
A_{\pi\pi}={\cal T}e^{i\delta_T}e^{i\gamma}+{\cal P}e^{i\delta_P}e^{-i\beta}~~,
$$
$$
A_{\pi K}=\tilde{r}_u{\cal T}e^{i\delta_T}e^{i\gamma}-r_t\tilde{\cal P}
e^{i\delta_P}~~,
$$
\beq \label{eqn:withbk}
A_+=r_t\tilde{\cal P}e^{i\delta_P}~~.
\eeq
It will be shown that the numerous {\it a priori} unknown parameters in
(\ref{eqn:withbk}), including the two weak phases
$\alpha\equiv\pi-\beta-\gamma$ and $\gamma$, can be
determined from the rate measurements of the above three processes and their
charge-conjugates.

First, we note that the amplitudes for the corresponding charge-conjugate decay
processes are simply obtained by changing the signs of the weak phases $\gamma$
and $\beta$. We denote the charge-conjugate amplitudes corresponding to (6) by
$\overline{A}_{\pi\pi},~\overline{A}_{\pi K},~A_-$, respectively.

The time-dependent tagged $B^0$ and $\overline{B}^0$ decay rates to
$\pi^+\pi^-$ are given by
$$
\Gamma(B^0(t)\to\pi^+\pi^-)=e^{-\Gamma t}[|A_{\pi\pi}|^2\cos^2({\Delta m\over
2}t)+|\overline{A}_{\pi\pi}|^2\sin^2({\Delta m\over 2}t)
$$
$$+{\rm Im}(e^{2i\beta}
A_{\pi\pi}\overline{A}^*_{\pi\pi})\sin(\Delta mt)]~~,
$$
$$
\Gamma(\overline{B}^0(t)\to\pi^+\pi^-)=e^{-\Gamma
t}[|A_{\pi\pi}|^2\sin^2({\Delta
m\over 2}t)+|\overline{A}_{\pi\pi}|^2\cos^2({\Delta m\over 2}t)
$$
\beq
-{\rm
Im}(e^{2i\beta} A_{\pi\pi}\overline{A}^*_{\pi\pi})\sin(\Delta mt)]~~.
\eeq
Measurement of these rates determines
$|A_{\pi\pi}|^2,~|\overline{A}_{\pi\pi}|^2$ and ${\rm Im}(e^{2i\beta}
A_{\pi\pi}\overline{A}^*_{\pi\pi})$:
$$
|A_{\pi\pi}|^2={\cal T}^2+{\cal P}^2-2{\cal T}{\cal P}\cos(\delta-\alpha)~~,
$$
$$
|\overline{A}_{\pi\pi}|^2={\cal T}^2+{\cal P}^2-2{\cal T}{\cal
P}\cos(\delta+\alpha)~~,
$$
\beq
{\rm Im}(e^{2i\beta} A_{\pi\pi}\overline{A}^*_{\pi\pi})=-{\cal
T}^2\sin(2\alpha)
+2{\cal T}{\cal P}\cos\delta\sin\alpha~~,
\eeq
where we used $\beta+\gamma=\pi-\alpha$ and where we defined
$\delta\equiv\delta_T-\delta_P$. The rates of the self-tagging modes $\pi^-
K^+,~\pi^+ K^-$ and $\pi^+K^0$ determine $|A_{\pi K}|^2,~|\overline{A}_{\pi
K}|^2$ and $|A_+|^2$, respectively:
$$
|A_{\pi K}|^2=\tilde{r}^2_u{\cal T}^2+r^2_t\tilde{\cal P}^2-2\tilde{r}_u
r_t{\cal T}\tilde{\cal P}\cos(\delta+\gamma)~~,
$$
$$
|\overline{A}_{\pi K}|^2=\tilde{r}^2_u{\cal T}^2+r^2_t\tilde{\cal P}^2-
2\tilde{r}_u r_t{\cal T}\tilde{\cal P}\cos(\delta-\gamma)~~,
$$
\beq
|A_+|^2=|A_-|^2=r^2_t\tilde{\cal P}^2~~.
\eeq

Measurement of the six quantitities in (8)$-$(9) suffices to determine all six
parameters $\alpha,~\gamma,~{\cal T},~{\cal P},~ \tilde{\cal P},~\delta$ up to
discrete ambiguities. The CKM parameter $r_t\equiv|V_{ts}/V_{td}|$, which is
still largely unknown, is obtained from the unitarity triangle in terms of
$\alpha$ and $\gamma$:
\beq
r_u r_t={\sin\alpha\over\sin\gamma}~~.
\eeq

We note immediately that
\beq
|A_{\pi K}|^2-|\overline{A}_{\pi K}|^2=
-({f_K\over f_{\pi}})({\tilde{\cal P}\over{\cal P}})(|A_{\pi\pi}|^2-|
\overline{A}_{\pi\pi}|^2)~~,
\eeq
which determines the magnitude of SU(3) breaking in the penguin amplitude,
$\tilde{\cal P}/{\cal P}$. The relation (11) between the particle-antiparticle
rate differences in $B\to\pi K$ and in $B\to\pi\pi$ was recently derived
\cite{DH} in the SU(3) limit, $f_K/f_{\pi}\to 1,~\tilde{\cal P}/{\cal P}\to 1$.
The authors assumed for SU(3) breaking a value $\tilde{\cal P}/{\cal
P}=f_K/f_{\pi}$ (based on factorization of penguin amplitudes) which is
questionable. In our approach this ratio is a free parameter to be determined
by experiment. We expect it to differ from one by up to 30$\%$.

A combined sample of the decays $B^0\to\pi^+\pi^-$ and $B^0\to\pi^- K^+$ has
already been observed \cite{piK} with a joint branching ratio of about $2\times
10^{-5}$. Equal mixtures of the two modes are likely, although confirmation of
this estimate awaits a better $\pi/K$ separation. A similar branching ratio is
expected for $B^+\to\pi^+ K^0$, where the efficiency of observing a $K^0$ by a
$K_S$ decay to two charged pions is $1/3$. Samples of hundreds of events in
each of these modes
(combining $B^+\to\pi^+ K^0$ and $B^-\to\pi^-\overline{K}^0$)
are expected to be obtained in future $e^+e^-$ colliders
operating at the $\Upsilon$(4S) resonance. The resulting statistical accuracy
of determining the weak phases $\alpha$ and $\gamma$ using the above method
thus is
expected to be at a level of ten percent. The theoretical uncertainty of the
method is at a similar level, involving the following corrections all of which
are of order a few percent:  A correction from an electroweak
penguin amplitude in $B^+\to\pi^+ K^0$, corrections due to $u$ and $c$ quarks
in the $B^0\to\pi^+\pi^-$ penguin amplitude, second-order SU(3) breaking in the
magnitudes of weak amplitudes, and first order SU(3) breaking in the (small)
strong phase of the penguin amplitude.

To summarize, we have shown that measurements of the rates for $B$ decays to
modes involving charged pions and kaons in the final states can determine the
shape of the unitarity triangle.
The accuracy of this method of determining the angles $\alpha$ and $\gamma$ in
future $e^+e^-$ $B$ factories is roughly estimated to be at a
level of 10$\%$. More detailed studies of the precision of this method are
worthwhile.

We thank J. Bjorken and H. Quinn for fruitful discussions, and
the Aspen Center for Physics for a congenial atmosphere in which the
first part of this collaboration was carried out. M. G. wishes to
acknowledge the hospitality of the SLAC theory group
during parts of this investigation. This work was supported in part by the
United States -- Israel Binational Science Foundation under Research Grant
Agreement 94-00253/1, by the Fund for Promotion of Research at the Technion,
and by the United States Department of Energy under Contract No. DE FG02
90ER40560.
\bigskip

\def \ajp#1#2#3{Am.~J.~Phys.~{\bf#1}, #2 (#3)}
\def \apny#1#2#3{Ann.~Phys.~(N.Y.) {\bf#1} (#3) #2}
\def \app#1#2#3{Acta Phys.~Polonica {\bf#1} (#3) #2 }
\def \arnps#1#2#3{Ann.~Rev.~Nucl.~Part.~Sci.~{\bf#1} (#3) #2}
\def \cmp#1#2#3{Commun.~Math.~Phys.~{\bf#1} (#3) #2}
\def \cmts#1#2#3{Comments on Nucl.~Part.~Phys.~{\bf#1} (#3) #2}
\def \cn{Collaboration}
\def \corn93{{\it Lepton and Photon Interactions:  XVI International Symposium,
Ithaca, NY August 1993}, AIP Conference Proceedings No.~302, ed.~by P. Drell
and D. Rubin (AIP, New York, 1994)}
\def \cp89{{\it CP Violation,} edited by C. Jarlskog (World Scientific,
Singapore, 1989)}
\def \dpff{{\it The Fermilab Meeting -- DPF 92} (7th Meeting of the American
Physical Society Division of Particles and Fields), 10--14 November 1992,
ed. by C. H. Albright \ite~(World Scientific, Singapore, 1993)}
\def \dpf94{DPF 94 Meeting, Albuquerque, NM, Aug.~2--6, 1994}
\def \efi{Enrico Fermi Institute Report No. EFI}
\def \el#1#2#3{Europhys.~Lett.~{\bf#1} (#3) #2}
\def \f79{{\it Proceedings of the 1979 International Symposium on Lepton and
Photon Interactions at High Energies,} Fermilab, August 23-29, 1979, ed.~by
T. B. W. Kirk and H. D. I. Abarbanel (Fermi National Accelerator Laboratory,
Batavia, IL, 1979}
\def \hb87{{\it Proceeding of the 1987 International Symposium on Lepton and
Photon Interactions at High Energies,} Hamburg, 1987, ed.~by W. Bartel
and R. R\"uckl (Nucl. Phys. B, Proc. Suppl., vol. 3) (North-Holland,
Amsterdam, 1988)}
\def \ib{{\it ibid.}~}
\def \ibj#1#2#3{~{\bf#1}, #2 (#3)}
\def \ichep72{{\it Proceedings of the XVI International Conference on High
Energy Physics}, Chicago and Batavia, Illinois, Sept. 6--13, 1972,
edited by J. D. Jackson, A. Roberts, and R. Donaldson (Fermilab, Batavia,
IL, 1972)}
\def \ijmpa#1#2#3{Int.~J.~Mod.~Phys.~A {\bf#1} (#3) #2}
\def \ite{{\it et al.}}
\def \jmp#1#2#3{J.~Math.~Phys.~{\bf#1} (#3) #2}
\def \jpg#1#2#3{J.~Phys.~G {\bf#1} (#3) #2}
\def \lkl87{{\it Selected Topics in Electroweak Interactions} (Proceedings of
the Second Lake Louise Institute on New Frontiers in Particle Physics, 15--21
February, 1987), edited by J. M. Cameron \ite~(World Scientific, Singapore,
1987)}
\def \ky85{{\it Proceedings of the International Symposium on Lepton and
Photon Interactions at High Energy,} Kyoto, Aug.~19-24, 1985, edited by M.
Konuma and K. Takahashi (Kyoto Univ., Kyoto, 1985)}
\def \mpla#1#2#3{Mod.~Phys.~Lett.~A {\bf#1} (#3) #2}
\def \nc#1#2#3{Nuovo Cim.~{\bf#1} (#3) #2}
\def \np#1#2#3{Nucl.~Phys.~B{\bf#1}, #2  (#3)}
\def \pisma#1#2#3#4{Pis'ma Zh.~Eksp.~Teor.~Fiz.~{\bf#1} (#3) #2[JETP Lett.
{\bf#1} (#3) #4]}
\def \pl#1#2#3{Phys.~Lett.~{\bf#1}, #2 (#3)}
\def \plb#1#2#3{Phys.~Lett.~B{\bf#1}, #2 (#3)}
\def \pra#1#2#3{Phys.~Rev.~A {\bf#1} (#3) #2}
\def \prd#1#2#3{Phys.~Rev.~D {\bf#1}, #2 (#3)}
\def \prl#1#2#3{Phys.~Rev.~Lett.~{\bf#1}, #2 (#3)}
\def \prp#1#2#3{Phys.~Rep.~{\bf#1} (#3) #2}
\def \ptp#1#2#3{Prog.~Theor.~Phys.~{\bf#1}, #2 (#3)}
\def \rmp#1#2#3{Rev.~Mod.~Phys.~{\bf#1} (#3) #2}
\def \rp#1{~~~~~\ldots\ldots{\rm rp~}{#1}~~~~~}
\def \si90{25th International Conference on High Energy Physics, Singapore,
Aug. 2-8, 1990}
\def \slc87{{\it Proceedings of the Salt Lake City Meeting} (Division of
Particles and Fields, American Physical Society, Salt Lake City, Utah, 1987),
ed.~by C. DeTar and J. S. Ball (World Scientific, Singapore, 1987)}
\def \slac89{{\it Proceedings of the XIVth International Symposium on
Lepton and Photon Interactions,} Stanford, California, 1989, edited by M.
Riordan (World Scientific, Singapore, 1990)}
\def \smass82{{\it Proceedings of the 1982 DPF Summer Study on Elementary
Particle Physics and Future Facilities}, Snowmass, Colorado, edited by R.
Donaldson, R. Gustafson, and F. Paige (World Scientific, Singapore, 1982)}
\def \smass90{{\it Research Directions for the Decade} (Proceedings of the
1990 Summer Study on High Energy Physics, June 25 -- July 13, Snowmass,
Colorado), edited by E. L. Berger (World Scientific, Singapore, 1992)}
\def \stone{{\it B Decays}, edited by S. Stone (World Scientific, Singapore,
1994)}
\def \tasi90{{\it Testing the Standard Model} (Proceedings of the 1990
Theoretical Advanced Study Institute in Elementary Particle Physics, Boulder,
Colorado, 3--27 June, 1990), edited by M. Cveti\v{c} and P. Langacker
(World Scientific, Singapore, 1991)}
\def \yaf#1#2#3#4{Yad.~Fiz.~{\bf#1} (#3) #2 [Sov.~J.~Nucl.~Phys.~{\bf #1} (#3)
#4]}
\def \zhetf#1#2#3#4#5#6{Zh.~Eksp.~Teor.~Fiz.~{\bf #1} (#3) #2 [Sov.~Phys. -
JETP {\bf #4} (#6) #5]}
\def \zpc#1#2#3{Zeit.~Phys.~C {\bf#1}, #2  (#3)}


\begin{thebibliography}{99}

\bibitem{CPreview}
For reviews, see, for example, Y. Nir and H. R. Quinn in \stone, p.\ 362;
I. Dunietz, {\it ibid.}, p.\ 393; M. Gronau, {\it Proceedings of Neutrino
94, XVI International Conference on Neutrino Physics and Astrophysics},
Eilat, Israel, May 29 -- June 3, 1994, eds. A. Dar, G. Eilam and M. Gronau,
Nucl.\ Phys.\ (Proc. Suppl.) B{\bf 38}, 136 (1995); J. L. Rosner, \efi
95-36, hep-ph/9506364, lectures presented at VIII J. A. Swieca, Summer School,
Rio de Janeiro, Feb.~7--11, 1995, proceedings to be published by World
Scientific.

\bibitem{CKM} N. Cabibbo, \prl{10}{531}{1963}: M. Kobayashi and T. Maskawa,
\ptp{49}{652}{1973}.

\bibitem{CCFT} J. H. Christenson, J. W. Cronin, V. L. Fitch, and R. Turlay,
\prl{13}{138}{1964}.

\bibitem {Penguin} M. Gronau, \prl {63}{1451}{1989};
D. London and R.D. Peccei, \plb{223}{257}{1989}; B. Grinstein, \plb{229}
{280}{1989}.

\bibitem{EWP} R. Fleischer, \zpc{62}{81}{1994}; \plb{321}{259}{1994};
\ibj{332}{419}{1994}; N. G. Deshpande and X.-G. He, \prl{74}{26}{1995}.

\bibitem{Isospin} M. Gronau and D. London, \prl{65}{3381}{1990}.

\bibitem{EWPin} M. Gronau, O. F. Hern\'andez, D. London, and J. L. Rosner,
Technion report TECHNION-PH-95-11, April, 1995, hep-ph/9504327, to be published
in \prd{52}{November 1}{1995}.

\bibitem{DS} F. DeJongh and P. Sphicas, Fermilab report FERMILAB-PUB-95/179-E,
July 1995.

\bibitem{SilWo} J. Silva and L. Wolfenstein, \prd{49}{R1151}{1994}.

\bibitem{BurFl} A. J. Buras and R. Fleischer, Technical University of
Munich report TUM-T31-96-95, July, 1995, hep-ph/9507460.

\bibitem{gamma} M. Gronau and D. Wyler, \plb{265}{172}{1991}; M. Gronau,
D. London, and J. L. Rosner, \prl{73}{21}{1994};
M. Gronau \ite, Ref. \cite{EWPin};
N. G. Deshpande and X.-G. He, University of Oregon preprint
OITS-576, May, 1995, hep-ph/9505369 (unpublished); M. Gronau and J. L. Rosner,
Technion report TECHNION-PH-95-26, September, 1995, hep-ph/9509325, submitted
to Phys.~Rev.~D.

\bibitem{DZ} D. Zeppenfeld, \zpc{8}{77}{1981}.

\bibitem{SW} M. Savage and M. Wise, \prd{39}{3346}{1989}; \ibj{40}{3127(E)}
{1989}.

\bibitem{Chau} L. L. Chau \ite, \prd{43}{2176}{1991}.

\bibitem{SU3} M. Gronau, O. F. Hern\'andez, D. London, and J. L. Rosner,
\prd{50}{4529}{1994}; O. F. Hern\'andez, D. London, M. Gronau, and J. L.
Rosner, \plb{333}{500}{1994}.

\bibitem{SU3br} M. Gronau, O. F. Hern\'andez, D. London, and J. L. Rosner,
Technion Report TECHNION-PH-95-10, March, 1995, hep-ph/9504326, to be published
in \prd{52}{November 1}{1995}.

\bibitem{BF} A. J. Buras and R. Fleischer, \plb{341}{379}{1995}.

\bibitem{MG} M. Gronau, \plb{300}{163}{1993}.

\bibitem{Browder} T. Browder, K. Honscheid and S. Playfer, in \stone,  p.~158.

\bibitem{Penphase} M. Bander, D. Silverman and A. Soni, \prl{43}{242}{1979};
G. Eilam, M. Gronau and J. L. Rosner, \prd{39}{819}{1989}; L. Wolfenstein,
\prd{43}{151}{1991} J.-M. G\'erard and W.-S. Hou, \prd{43}{2909}{1991};
H. Simma, G. Eilam and D. Wyler, \np{352}{367}{1991}.

\bibitem{DH} N. G. Deshpande and X.-G. He, \prl{75}{1703}{1995}.

\bibitem{piK} M. Battle {\it et al.} (CLEO Collaboration),
\prl{71}{3922}{1993}.

\end{thebibliography}
\end{document}